\documentstyle[12pt]{article}
\def\GeV{\,{\rm GeV}}

\def\sec{\,{\rm sec}}
\def\Gyr{\,{\rm Gyr}}

\def\rcm{\,{\rm cm}}

\def\Mpc{\,{\rm Mpc}}

\def\eV{{\,\rm eV}}

\def\cmm2{{\,\rm cm^{-2}}}
\def\cm2{{\,{\rm cm}^2}}
\def\cmm3{{\,{\rm cm}^{-3}}}
\def\gcmm3{{\,{\rm g\,cm^{-3}}}}
\def\kms{\,{\rm km\,s^{-1}}}

\def\la{\mathrel{\mathpalette\fun <}}

\def\fun#1#2{\lower3.6pt\vbox{\baselineskip0pt\lineskip.9pt
  \ialign{$\mathsurround=0pt#1\hfil##\hfil$\crcr#2\crcr\sim\crcr}}}

\begin{document}
\pagestyle{empty}
\begin{center}
\bigskip

\rightline{FERMILAB--Pub--96/065-A}
\rightline{astro-ph/9603081}
\rightline{submitted to {\it Science}}

\vspace{.2in}
{\Large \bf Cold Dark Matter Models}
\bigskip

\vspace{.2in}
Scott Dodelson,$^1$ Evalyn I.~Gates,$^{1,2}$ and
Michael S. Turner$^{1,2,3}$\\

\vspace{.2in}
{\it $^1$NASA/Fermilab Astrophysics Center\\
Fermi National Accelerator Laboratory, Batavia, IL~~60510-0500}\\

\vspace{0.1in}
{\it $^2$Department of Astronomy \& Astrophysics\\
Enrico Fermi Institute, The University of Chicago, Chicago, IL~~60637-1433}\\

\vspace{0.1in}
{\it $^3$Department of Physics\\
Enrico Fermi Institute, The University of Chicago, Chicago, IL~~60637-1433}\\

\end{center}

\vspace{.3in}
\centerline{\bf ABSTRACT}
\bigskip
Motivated by inflation, the theory of big-bang nucleosynthesis
and the quest for a deeper understanding of the fundamental
forces and particles, a very successful paradigm
for the development of structure in the Universe has evolved.
It holds that most of
the matter exists in the form of slowly moving
elementary particles left over from the earliest moments (cold
dark matter or CDM) and that the small density inhomogeneities that
seed structure formation arose from quantum fluctuations around
$10^{-34}\sec$ after the bang.  A flood of
observations is now testing the cold dark matter paradigm -- from
determinations of the Hubble constant to measurements of
the anisotropy of the Cosmic Background Radiation (CBR) -- and
could reveal the details of the theory as well.

\newpage
\pagestyle{plain}
\setcounter{page}{1}
\newpage

\section*{Introduction}

According to the highly successful hot big-bang model
the Universe began as a hot, smooth soup of the fundamental
particles \cite{bigbang}.   On the other hand, the most conspicuous feature
of the Universe today is the
abundance of structure -- stars, galaxies, clusters of
galaxies, superclusters, voids, and very large, sheet-like
structures comprised of galaxies (dubbed great walls) \cite{structure}.
The extreme uniformity of the temperature of the Cosmic Background
Radiation (average temperature $2.728\,{\rm K}\pm 0.002\,$K.)
indicates that this
structure must have arisen from very small, primeval inhomogeneities in
the distribution of matter.  It is believed this occurred
through the attractive action
of gravity operating over the past 15 Gyr, amplifying
the primeval inhomogeneities by a factor of more than $10^5$ \cite{gravinstab}.

This general picture was confirmed in 1992 when the Differential
Microwave Radiometer (DMR) on NASA's Cosmic Background
Explorer (COBE) satellite
detected differences in the CBR temperature in directions separated
on the sky by $10^\circ$ at the level of $30\mu$K ($\delta
T/T \approx 10^{-5}$) \cite{dmr}, providing the first
evidence for the existence of the matter inhomogeneity that
seeded all structure.  Since then CBR anisotropy
of a similar size has been detected by more than ten experiments
on angular scales from $0.5^\circ$ to tens of degrees (see Fig.~1)
\cite{cbrsummary}.
The challenge is to put together a detailed and coherent picture
of structure formation.  In doing so, many cosmologists believe that
much will be revealed about the earliest moments of the
Universe and perhaps even the nature of the fundamental forces.

Several approaches to structure formation have been pursued \cite{defect,pib};
the most successful is known as cold dark matter.
Its two basic tenets are:  (i) The Universe is spatially flat,
corresponding to a mean matter density equal to the critical
density, with ordinary matter (baryons) contributing about 5\% of the
critical density and slowly moving elementary particles left
over from the earliest moments (cold dark matter) contributing the rest;
and (ii) The primeval density perturbations are nearly
scale invariant
and arose from quantum-mechanical fluctuations occurring during
the earliest moments.  Scale-invariant refers to the fact that
fluctuations in the gravitational potential are independent of
length scale.  More precisely, the Fourier components
of the primeval density field are drawn from
a gaussian distribution with variance given by
power spectrum $P(k) \equiv \langle |\delta_k|^2
\rangle = Ak^n$ with $n\approx 1$, where $k=2\pi /\lambda$ is the
wavenumber.  For reference, perturbations of wavelength
around $1\Mpc$ gave rise to galaxies, $10\Mpc$ to clusters,
and $100\Mpc$ to the largest structures observed, where
$1\Mpc = 3.09\times 10^{25}\rcm$.

Cold dark matter draws from three important ideas --
inflation, big-bang nucleosynthesis, and the quest to
better understand the fundamental forces and particles.  Inflation holds that
the
Universe underwent a very early ($t\sim 10^{-34}\sec$), very rapid period
of expansion during which it grew in size by a factor greater than
it has since.  This rapid expansion is driven by vacuum energy,
an  unusual form of energy predicted to exist by many theories which
unify the fundamental forces \cite{inflation}.  The enormous growth in size
leads to a Universe which
appears to be flat on the length scales that we can probe (up
to the current horizon, $d_H \sim 3000 h^{-1}\Mpc$), and thus
has critical density.   Further, the tremendous growth
allows quantum-mechanical
fluctuations excited on extremely small scales ($\ll 10^{-16}\rcm$)
to be stretched in length to become variations in the energy
density on astrophysical scales.  (The conversion from quantum
fluctuations to energy fluctuations occurs when the vacuum energy
decays into radiation at the end of inflation.)

Big-bang nucleosynthesis refers to the very successful description
of how the light-elements D, $^3$He, $^4$He and $^7$Li were produced by
nuclear reactions during the first few seconds \cite{bbn}.
The agreement of the predicted
and measured light-element abundances is an important confirmation
of the hot big-bang model and leads to the most precise
determination of the density of ordinary matter.  The baryon
density inferred from nucleosynthesis is between $1.7\times 10^{-31}
\gcmm3$ and $4.1 \times 10^{-31}\gcmm3$
and corresponds to a fraction of critical
density which depends upon the value of the Hubble constant,
$\Omega_B = 0.01h^{-2} - 0.02h^{-2}$, where
$H_0 = 100h \kms \Mpc^{-1}$ \cite{bbn}.  Allowing $h=0.4 -0.9$,
 consistent with modern measurements \cite{h0rev},
implies that ordinary matter can contribute at most 15\% of
the critical density.  If the inflationary prediction
is correct, then most of the matter in the Universe must be
nonbaryonic (see Fig.~2).

This idea has received indirect support from particle physics.
Attempts to further our understanding of the particles
and forces have led to the prediction
of new, stable or long-lived particles that interact very feebly with ordinary
matter.  These particles, if they exist, should have
been present in great numbers during the earliest moments
and remain today in numbers sufficient to contribute
the critical density \cite{pdm}.  Two of the most attractive
possibilities behave like cold dark matter:  a neutralino of mass
$10\GeV$ to $1000\GeV$, predicted in supersymmetric theories, and an axion
of mass $10^{-6}\eV$ to $10^{-4}\eV$, needed to solve a subtle problem of the
standard model of particle physics (strong-CP problem).
The third interesting possibility
is that one of the three neutrino species has a mass between $5\eV$ and
$30\eV$; neutrinos move very fast and are referred to
as hot dark matter.\footnote{The possibility that most of the exotic
particles are fast-moving neutrinos -- hot dark matter -- was explored first
and found to be inconsistent with observations \cite{hdm}.  The
problem is that large structures form first and must fragment into
smaller structures, which conflicts with the fact that
large structures are just forming today.}

According to cold dark matter theory CDM particles
provide the cosmic infrastructure:  It is their gravitational
attraction that forms and holds cosmic structures together.
Structure forms in a hierarchical manner, with galaxies forming
first and successively larger objects forming thereafter \cite{faberetal}.
Quasars and other rare objects form at redshifts of up to
five, with ordinary galaxies forming a short time later.  Today, superclusters,
objects made of several
clusters of galaxies, are just becoming bound by the gravity of
their CDM constituents.  The formation of larger and larger objects
continues.  In the clustering process regions of space are left
devoid of matter -- and galaxies -- leading to voids.
If the CDM theory is correct,  CDM particles
are the ubiquitous dark matter known only by its gravitational
effects which accounts for most of
the mass density in the Universe and holds galaxies,
clusters of galaxies and even the Universe itself together \cite{darkmatter}.

\section*{Standard Cold Dark Matter}

When the cold dark matter scenario emerged more than
a decade ago many referred to it as a no parameter
theory because it was so specific compared to previous
models for the formation of structure.  This was an overstatement
as there are cosmological quantities that must be
known to determine the development of structure in detail.
However, the data available did not require precise knowledge of
these quantities to test the model.

Broadly speaking these parameters can be organized
into two groups.  First are the cosmological parameters:  the
Hubble constant, $H_0=100h \kms\Mpc^{-1}$; the density of
ordinary matter, specified by $\Omega_B h^2$; the power-law index $n$ and
normalization $A$ that quantify the density perturbations.\footnote{In
addition, inflation also predicts a nearly scale-invariant
spectrum of gravitational waves, which can also lead to
CBR anisotropy that is not simple to distinguish from that
caused by density perturbations.   If the normalization $A$ is
determined from CBR anisotropy measurements, as is now
done, the level of gravitational radiation must be specified
(denoted by $T/S$ for tensor to scalar ratio) as well as the
power-law index for gravity waves, $n_T$.}
The inflationary parameters fall into this category
because there is no standard model of inflation.
On the other hand, once determined they can be used to discriminate
between models of inflation.

The other quantities specify the composition of invisible matter
in the Universe:  radiation, dark matter, and a possible cosmological
constant.  Radiation refers to relativistic
particles:  the photons in the CBR, three massless
neutrino species (assuming none of the neutrino species has
a mass), and possibly other undetected relativistic particles
(some particle-physics theories predict the existence of additional
massless particle species).  At present relativistic particles
contribute almost nothing to the energy density in the Universe,
$\Omega_R \simeq 4.2 \times 10^{-5}h^{-2}$; early on -- when
the Universe was smaller than about $10^{-5}$ of its present
size -- they dominated the energy content.

In addition to CDM particles, the dark matter could include other particle
relics.  For example, each neutrino species has a number density of
$113\cmm3$, and a neutrino species of mass $5\eV$
would account for about 20\% of the critical density
($\Omega_\nu = m_\nu/90h^{2}\eV$).  Predictions for
neutrino masses range from $10^{-12}\eV$ to several MeV, and
there is some experimental evidence that at least one of
the neutrino species has a small mass \cite{numass}.

Finally, there is the cosmological constant.  Both introduced
and abandoned by Einstein, it is still with us.
In the modern context it corresponds
to an energy density associated with the quantum vacuum.  At present,
there is no reliable calculation of the value that the cosmological
constant should take, and so its existence
must be regarded as a logical possibility.

The original no parameter cold dark matter model,
referred to as standard CDM, is characterized by:  $h=0.5$, $\Omega_B =0.05$,
$\Omega_{\rm CDM}=0.95$, $n=1$, and
standard radiation content.  The overall normalization of the
density perturbations was fixed by
comparing the predicted level of inhomogeneity with that
seen today in the distribution of bright galaxies.
Specifically, the amplitude $A$ was
determined by comparing the expected mass fluctuations in
spheres of radius $8h^{-1}\Mpc$ (denoted by $\sigma_8$) to the
galaxy-number fluctuations in spheres of the same size.
The galaxy-number fluctuations on the scale $8h^{-1}\Mpc$
are unity; adjusting $A$ to achieve $\sigma_8 = 1$ corresponds
to the assumption that light, in the form of bright galaxies, traces mass.
Choosing $\sigma_8$ to be less than one means that light is more
clustered than mass and is a biased tracer of mass.
There is some evidence that bright galaxies are somewhat
more clumped than mass with biasing factor $b\equiv 1/\sigma_8
\simeq 1 - 2$ \cite{biasing}.

A dramatic change occurred with the detection of CBR anisotropy
by COBE in 1992 \cite{dmr}.
The COBE measurement permitted a precise normalization of the
amplitude of density perturbations on very large scales
($\lambda \sim 10^4h^{-1}\Mpc$) without regard to the issue of biasing.
[CBR anisotropy on the angular scale $\theta$ arises primarily due to
inhomogeneity on length scales $\lambda \sim 100h^{-1}
\Mpc (\theta /{\rm deg})$.]
For standard CDM, the COBE normalization leads to:  $\sigma_8
=1.2 \pm 0.1 $ or anti-bias since $b=1/\sigma_8\simeq 0.7$.  The
pre-COBE normalization ($\sigma_8 = 0.5$) led to too little power
on scales of $30h^{-1}\Mpc$ to $300h^{-1}\Mpc$, as compared
to what was indicated in redshift surveys, the angular correlations
of galaxies on the sky and the peculiar velocities of galaxies.
The COBE normalization leads to about the right amount of power on
these scales, but appears to predict too much power on small
scales ($\la 8 h^{-1}\Mpc$).
While standard CDM is in general agreement with the observations, a
consensus has developed that the conflict just mentioned is probably
significant \cite{jpo-ll} -- and we concur.  This has led to a new look at
the cosmological and invisible-matter parameters and
to the realization that the problems of standard CDM
are simply a poor choice for the standard parameters.

\section*{The CDM Family of Models}

Standard CDM has served well as an industry-wide standard
that focused everyone's attention --
the DOS of cosmology if you will.
However, the quality and quantity of data have improved
and knowledge of the cosmological and invisible-matter parameters
has become important for serious testing of CDM and inflation.
As we shall discuss, there are a variety of
combinations of the parameters that
lead to good agreement with the existing data on both large
and small length scales -- and thus can make a claim to being
the new standard CDM model.  To illustrate, we have
compared COBE-normalized CDM models with measurements
of the distribution of matter in the Universe and in Figs.~4-7
show the allowed values of the cosmological and
invisible-matter parameters.

More precisely, for a given CDM model -- specified by
the cosmological and invisible-matter parameters --
we compute the expected CBR anisotropy and require that
it be consistent with the four-year COBE data set at
the two-sigma level \cite{dmr4yr}.\footnote{The predicted COBE anisotropy is
relatively insensitive to all the parameters except
$A$, $n$ and the level of gravitational radiation;
for a given level of gravitational radiation and $n$,
COBE fixes $A$.}  Having COBE-normalized the model we compute the expected
level
of inhomogeneity in the Universe today and compare to three robust
measurements of inhomogeneity.\footnote{Computation of both the CBR
anisotropy and the level of inhomogeneity
today depends upon the invisible-matter content and the cosmological
parameters and requires that the distribution of
matter and radiation be evolved numerically.
We have carried out these calculations by
integrating the coupled Einstein and Boltzmann equations; for
details see Ref.~\cite{boltzmann}.}

The first measurement is the shape of the power spectrum as
inferred from surveys of the distribution of galaxies today.
(Because the distance to a given galaxy is determined from its
redshift through the Hubble law, $d = zH_0$, such surveys are called
redshift surveys.)   We have used the analysis of Peacock and Dodds \cite{pd}
(see Fig.~3).
In the absence of an understanding
of the relationship between the distributions of
light and mass we leave the bias factor as a free parameter.

The next measurement is a determination of $\sigma_8$.
The abundance of rich, x-ray emitting clusters is sensitive to the
level of inhomogeneity on scales around $8h^{-1}\Mpc$ and
thus provides a good means of inferring the value of $\sigma_8$.
Following Efstathiou, White and
Frenk \cite{sigma8cluster} we use $0.5 \le \sigma_8 \le 0.8 $ for
models with $\Omega_{\rm Matter} = 1$ and
let this range scale with $\Omega_{\rm Matter}^{-0.56}$
for models with a cosmological constant ($\Omega_{\Lambda} = 1 -
\Omega_{\rm Matter}$).

The final observational constraint involves the formation of
objects at high redshift (early structure formation).  At
redshifts of two to four, hydrogen clouds, detected by their
absorption features in the spectra of high-redshift
quasars ($z\sim 4 - 5$), contribute a fraction of
the critical density, $\Omega_{\rm clouds} \simeq (0.001\pm 0.0002)h^{-1}$
\cite{Dlya}.  It is believed that these objects are the forerunners of
bright galaxies.  Insisting that the predicted level of inhomogeneity
is sufficient to account for the number of these protogalaxies
seen at redshift four leads to a lower limit to the power
on very small scales ($\lambda \sim 0.2h^{-1}\Mpc$).\footnote{Following others,
we
have used the Press-Schechter formalism to compute
the number of clouds that form by a given redshift \cite{clouds}.}

Figure 4 gives the overall picture.  The simplest CDM models --
those with standard invisible-matter content -- lie in a
region that runs diagonally from smaller Hubble constant
and larger $n$ to larger Hubble constant and smaller $n$.  That is, higher
values of the Hubble constant require more tilt (tilt referring
to deviation from scale invariance).
Note too that standard CDM is well outside of the allowed range.
Current measurements of CBR anisotropy on the degree scale,
as well as the COBE four-year anisotropy data,
preclude $n$ less than about 0.7 (see Fig.~1).  This implies that the
largest Hubble constant consistent with the simplest CDM
models is slightly less than $60\kms\Mpc^{-1}$.
If the invisible-matter content is nonstandard, higher values of the
Hubble constant can be accommodated.

Figure 5 illustrates the interplay of $\Omega_\nu$, $h$ and $n$.
Higher values of the Hubble constant require more tilt and/or higher
neutrino content.  There are two interesting things to notice.
Even a neutrino content as low as 5\% (corresponding to a neutrino mass
of around $1\eV$ or so) has significant consequence -- it allows
very nearly scale invariant perturbations ($n> 0.9$).
The jagged shape of the allowed region in Fig.~4 arises because
$\Omega_\nu =0.2$ is almost precluded for any value of $h$ and $n$
and any higher content of hot dark matter is not viable at all
\cite{nucontent}.

Figure 6 illustrates how the introduction of a cosmological
constant can easily accommodate larger values of the Hubble constant,
and Fig.~7 shows the close relationship between the
Hubble constant and the radiation content.  The ratio between
the matter density and the radiation density determines the shape
of the power spectrum -- so that lowering $h$ or raising $g_*$
have precisely the same effect.  ($g_* = 2 +0.454N_\nu$
quantifies the energy density in relativistic particles,
where $N_\nu$ is the equivalent number of neutrino species.)

Changes in the different parameters from their standard CDM values
alleviate the excess power on small scales in different ways.
Tilt has the effect of reducing
power on small scales when power on very large scales is fixed by COBE.
A small admixture of hot dark matter works because
fast moving neutrinos suppress the growth
of inhomogeneity on small scales by streaming from
regions of higher density and to regions of lower density.
(It was in fact this feature of hot dark matter that led to
the demise of the hot dark matter model for structure formation.)

A low value of the Hubble constant, additional radiation
or a cosmological constant all reduce power on small scales by
lowering the ratio of matter to radiation.  Since the
critical density depends upon the square of the
Hubble constant, $\rho_{\rm Critical} = 3H_0^2/8\pi G$,
a smaller value corresponds to a lower matter density
since $\rho_{\rm Matter} = \rho_{\rm Critical}$ for a flat
Universe without a cosmological constant.
Shifting some of the critical density to vacuum energy
also reduces the matter density since $\Omega_{\rm Matter} = 1 -
\Omega_\Lambda$.  Lowering the ratio of matter to radiation
reduces the power on small scales in a subtle way.  While the primeval
fluctuations in the gravitational potential are nearly scale-invariant,
density perturbations today are not because the Universe
made a transition from an early radiation-dominated phase ($t\la
1000\,$yrs), where the growth of density perturbations is inhibited,
to the matter-dominated phase, where growth proceeds unimpeded.
This introduces a feature in the power spectrum today
(see Fig.~3), whose location depends upon the relative amounts
of matter and radiation.  Lowering the ratio of matter
to radiation shifts the feature to larger scales and with power
on large scales fixed by COBE this leads to less power on small scales.

Some of the viable models have been discussed previously
as singular solutions
-- cosmological constant \cite{lcdm}, very low Hubble constant \cite{lhccdm},
tilt \cite{tcdm}, tilt + low Hubble constant \cite{stp},
extra radiation \cite{taucdm},
an admixture of hot dark matter \cite{nucdm}.  We wish to
emphasize that there is actually a continuum of viable models,
as can be seen in Figs.~4-7,
which arises because of imprecise knowledge of cosmological
parameters and the invisible-matter sector and not the
inventiveness of theorists.

\section*{Other Considerations}

There are many other observations that bear on structure
formation.  However, with cosmological data
systematic error and interpretational
issues are important considerations.  In fact,
if all extant observations
were taken at face value, there is no viable model for
structure formation, cold dark matter or otherwise!
With this as a preface, we now discuss some of the
other existing data as well as future measurements that
will more sharply test cold dark matter.

There is a general cosmological tension between measures of the age of the
Universe and determinations of the Hubble constant \cite{tension}.
It arises because determinations of the ages of the oldest stars lie
between $13\Gyr$ and $19\Gyr$ \cite{age} and recent measurements of the Hubble
constant favor values between $60\kms\Mpc^{-1}$ and
$80\kms\Mpc^{-1}$ \cite{h_0}, which, for $\Omega_{\rm Matter} =1$,
leads to a time back to the bang of $11\Gyr$ or less (see Fig.~8).\footnote{The
time back to the bang depends upon $H_0$, $\Omega_{\rm Matter}$
and $\Omega_\Lambda$; for $\Omega_{\rm Matter}=1$ and $\Omega_\Lambda
=0$, $t_{\rm BB}
= {2\over 3}H_0^{-1}$, or $13\Gyr$ for $h=0.5$ and $10\Gyr$
for $h=0.65$.  For a flat Universe with a cosmological constant
the numerical factor is larger than 2/3 (see Fig.~8).}
These age determinations receive additional support from
estimates of the age of the galaxy based upon the decay of
long-lived radioactive isotopes and the cooling of white-dwarf
stars, and all methods taken together make a strong case for
an absolute minimum age of $10\Gyr$ \cite{agereview}.
Within the uncertainties  there is no inconsistency,
though there is certainly tension, especially for $\Omega_{\rm Matter} =1$.

While age is not a major issue for cold dark matter -- large-scale
structure favors an older Universe by virtue of a lower Hubble
constant or cosmological constant -- the Hubble constant still
has great leverage.  If it is determined to be greater than about
$60\kms\Mpc^{-1}$, then only CDM models with nonstandard
invisible-matter content -- a cosmological constant or additional radiation
-- can be consistent with large-scale structure.  If
$H_0$ is greater than $65\kms\Mpc^{-1}$, consideration
of the age of the Universe
leaves $\Lambda$CDM as the lone possibility.  The issue
of $H_0$ is not settled, but the use of
Type Ia supernovae as standard candles, the study of
Cepheid variable stars in Virgo cluster galaxies using the Hubble
Space Telescope, and other methods make it likely that it will be soon.

If CDM is correct, baryons make up a small fraction of matter
in the Universe.  Most of the baryons in galaxy clusters are in the hot, x-ray
emitting intracluster gas and not the luminous galaxies.
The measured x-ray flux fixes the mass in baryons, while
the measured x-ray temperature fixes
the total mass (through the virial theorem).
The baryon-to-total-mass has been determined from x-ray measurements
for more than ten clusters and is found to be
$M_{\rm B}/M_{\rm TOT} \simeq (0.04-0.1)
h^{-3/2}$ \cite{gasratio}.  Because of their size,
clusters should represent a fair sample
of the cosmos and thus the baryon-to-total mass ratio should
reflect its universal value, $\Omega_B/\Omega_{\rm Matter}
\simeq (0.01 - 0.02)h^{-2}/\Omega_{\rm Matter}$.  These
two ratios are consistent for models with a very low
Hubble constant, $h\sim 0.4$ and $\Omega_{\rm Matter} = 1$,
or with a cosmological constant and $\Omega_{\rm Matter}\sim 0.3$.
However, important assumptions are made in this analysis
-- that the hot gas is unclumped and in virial
equilibrium and that magnetic fields do not provide significant
pressure support for the gas -- if any one of them is not valid the actual
baryon fraction would be smaller \cite{outs},\footnote{In fact,
there are some indications that cluster masses determined by the
weak-gravitational lensing technique lead to larger
values than the x-ray determinations \cite{weakgrav}.} allowing
for consistency with a larger value of $H_0$ without recourse
to a cosmological constant.

The halos of individual spiral galaxies like our own are not
large enough to provide a fair sample of matter in the
Universe -- for example, much
of the baryonic matter has undergone dissipation and condensed
into the disk of the galaxy.  Nonetheless, the content
of halos is expected to be primarily CDM particles.  This is
consistent with the fact that visible stars, hot gas, dust,
and even dark stars acting as microlenses (known as
MACHOs) account for only a fraction of the mass of our own halo
\cite{macho,nostars}.

Determining the mean mass density of the Universe
would discriminate between models with and without a cosmological constant,
as well as test the inflationary prediction of a flat
Universe.  A definitive determination is still lacking.
The measurement that averages over the largest volume --
and thus is potentially most useful -- uses
peculiar velocities of galaxies.  Peculiar velocities arise due to the
inhomogeneous distribution of matter, and the mean matter
density can be determined by relating the peculiar
velocities to the observed distribution of galaxies.  The results of this
technique
indicate that $\Omega_{\rm Matter}$ is at least 0.3 and perhaps as
large as unity \cite{strauss,dekel}.   Though not definitive,
this provides strong evidence for the existence of nonbaryonic dark
matter (see Fig.~2), a key aspect of cold dark matter.

A different approach to the mean density is through the deceleration
parameter $q_0$, which quantifies the slowing of the expansion
due to the gravitational attraction of matter in the Universe.
Its value is given by $q_0
 = {1\over 2}\Omega_{\rm Matter} - \Omega_\Lambda$ (vacuum energy actually
leads to accelerated expansion) and can be determined by relating the
distances and redshifts of distant objects.  In all but the $\Lambda$CDM
scenario, $q_0 =0.5$; for $\Lambda$CDM, $q_0 \sim -0.5$.
Two groups are trying to measure $q_0$ by using high redshift
($z\sim 0.7$) Type Ia supernovae as standard candles;
the preliminary results of one group suggest that $q_0$ is positive \cite{lbl}.
More than a dozen distant Type Ia supernovae were discovered
this year and both groups should soon have
enough to measure $q_0$ with a precision of $\pm 0.2$.

Gravitational lensing of distant QSOs by intervening galaxies
is another way to measure $q_0$, and the frequency of
QSO lensing suggests that $q_0 > -0.6$ \cite{qsolensing}.
The distance to a QSO of given redshift is larger
for smaller $q_0$, and thus the probability for its being
lensed by an intervening galaxy is greater.

The 10\,m Keck Telescope and the Hubble Space Telescope are
providing the deepest images of the Universe ever
and are revealing
details of galaxy formation as well as the formation and
evolution of clusters of galaxies.  The Keck has made the
first detection of deuterium in high redshift hydrogen
clouds \cite{D}.  This is a new confirmation of
big-bang nucleosynthesis and has the potential of pinning
down the density of ordinary matter to a precision of 10\%.

The level of inhomogeneity in the Universe today is determined largely
from redshift surveys, the largest of which contain of order $10^4$ galaxies.
A larger -- a million galaxy redshifts -- and more homogeneous
survey, the Sloan Digital Sky Survey, is in progress \cite{sdss}.  It
will allow the power spectrum to be measured more precisely
and out to large enough scales ($500h^{-1}\Mpc$)
to connect with measurements from CBR anisotropy on angular scales
of up to five degrees.

The most fundamental element of cold dark matter --
the existence of the CDM particles themselves -- is being tested.
While the interaction of CDM particles with ordinary matter occurs
through very feeble forces and makes their existence difficult to test,
experiments with sufficient sensitivity to
detect the CDM particles that hold our own galaxy together if they are
in the form of axions of mass $10^{-6}\eV - 10^{-4}\eV$ \cite{llnl} or
neutralinos of mass tens of GeV \cite{neutralinos} are now underway.
Evidence for the existence of the neutralino could also come
from particle accelerators searching for other supersymmetric particles.
In addition, several experiments sensitive to neutrino masses
are operating or are planned, ranging
from accelerator-based neutrino oscillation experiments
to the detection of solar neutrinos to the study of the
tau neutrino at $e^+e^-$ colliders.

CBR anisotropy probes the power spectrum most cleanly as
it is related directly to the distribution of matter
when density perturbations were very small \cite{cbranisotropy}.
Current measurements are beginning to test CDM and
differentiate between the variants (see Fig.~1); e.g., a spectral
index $n<0.7$ is strongly disfavored.  More than ten groups
are making measurements with instruments in space, on balloons and
at the South Pole.  Proposals have been made -- three
to NASA and one to ESA -- for a satellite-borne experiment in the
year 2000 that would map CBR anisotropy over the full sky
with $0.2^\circ$ resolution, about 30 times better than COBE.
The results from such a map could easily
discriminate between the different variants of CDM (see Fig.~9).

The first and most powerful test to emerge from these measurements
will be the location of the first (Doppler) peak in the angular power
spectrum (see Fig. 9).  All
variants of CDM predict the location of the first peak to lie
in roughly the same place. On the other hand, in an open Universe
(total energy density less than critical)
the first peak occurs at a larger value of $l$ (much smaller
angular scale).  This will provide an important test of inflation.
In addition,
theoretical studies \cite{learn} indicate that $n$ could be
determined to a precision of a few percent, $\Omega_\Lambda$
to ten percent, and perhaps even $\Omega_\nu$ to enough
precision to test $\nu$CDM \cite{US}.
Measurements of CBR anisotropy can also be used
to infer the value of the
inflationary potential and its first two derivatives \cite{recon}, which
could provide insight about the unification of
the forces and particles of Nature.

If {\it all} the current observations -- from recent Hubble
constant determinations to the cluster baryon fraction -- are taken at
face value, the cosmological constant + cold dark matter model
is probably the best fit \cite{bestfit}, though there may soon be a conflict
with the measurement of $q_0$ with Type Ia supernovae.
It raises a fundamental question -- the
origin of the implied vacuum energy, about $(10^{-2}\eV)^4$ --
since there is no known principle or mechanism that explains why
it is less than $(300\GeV )^4$, let alone $(10^{-2}\eV )^4$
\cite{cosmoconst}.  In any case, it would be imprudent to
take all the observational data at face value because of important
systematic and interpretational uncertainties.  To paraphrase the
biologist Francis Crick, a theory that fits all the data at
any given time is probably wrong as some of the data are
probably not correct.

\section*{Summary}

Cold dark matter is a bold attempt to extend
our knowledge of the Universe to within $10^{-32}\sec$
of the bang.  Current measurements and observations
are generally consistent with the cold dark matter theory, but because of
imprecise knowledge of important cosmological parameters as
well as the invisible-matter content of the Universe there are actually
a family of viable CDM models.

The number of observations that are testing the cold dark matter theory
is growing fast, and the prospects for discriminating between the
different variants in the next five years are excellent.
If cold dark matter is shown to be correct, an important aspect of
the standard cosmology -- the origin and evolution of structure --
will have been resolved and a window to the early
moments of the Universe and physics at very high energies will have been
opened.

\paragraph{Acknowledgments.}  This work was supported in part by
the DOE (at Chicago and Fermilab) and the NASA (at Fermilab through
grant NAG 5-2788).  We thank Martin White and other participants
of the Aspen Center for Physics workshop, {\it Inflation:
 From Theory to Observation and Back}, for many lively and useful discussions.

\newpage

\section*{Figure Captions}

\bigskip
\noindent{\bf Figure 1:}  Summary of CBR anisotropy measurements
and predictions for three CDM models (adapted from Ref.~\cite{cbrsummary}).
Plotted are the squares of the measured multipole amplitudes ($C_l = \langle
|a_{lm}|^2\rangle$) vs. multipole number $l$.
The temperature difference on angular scale $\theta$ is
given roughly by $\sqrt{l(l+1)|a_{lm}|^2}$ with $l\sim 200^\circ /\theta$.
The theoretical curves are standard CDM and CDM with $n=0.7$ and $h=0.5$.

\medskip
\noindent{\bf Figure 2:}  Summary of knowledge of $\Omega$.
The lowest band is luminous matter, in the form of bright
stars and related material; the middle band is the big-bang
nucleosynthesis determination of the density of baryons;
the upper region is the estimate of $\Omega_{\rm Matter}$
based upon the peculiar velocities of galaxies.  The
gaps between the bands illustrate the two dark matter problems:
most of the ordinary matter is dark and most of the matter is nonbaryonic.

\medskip
\noindent{\bf Figure 3:}  Measurements of the power spectrum,
$P(k) = |\delta_k|^2$, and the predictions of different
COBE-normalized CDM models.
The points are from several redshift surveys as analyzed
by Dodds and Peacock \cite{pd}; the models are:  $\Lambda$CDM
with $\Omega_\Lambda =0.6$ and $h=0.65$; standard CDM (sCDM),
CDM with $h=0.35$; $\tau$CDM (with the energy equivalent
of 12 massless neutrino species) and $\nu$CDM with $\Omega_\nu = 0.2$
(unspecified parameters have their standard CDM values).
The offset between the different models and the points
indicates the level of biasing implied.

\medskip
\noindent{\bf Figure 4:}  Acceptable values of the cosmological
parameters $n$ and $h$ for CDM models with standard
invisible-matter content (CDM), with 20\% hot dark matter ($\nu$CDM),
with additional relativistic particles (the energy equivalent
of 12 massless neutrino species, denoted $\tau$CDM), and with a cosmological
constant that accounts for 60\% of the critical density ($\Lambda$CDM).
Note that standard CDM ($n=1$ and $h=0.5$) is not viable.

\medskip
\noindent{\bf Figure 5:}  $\nu$CDM models:  Acceptable values of $\Omega_\nu$
and $h$ for $n=0.8, 0.9, 1.0$.  Note that no model is viable
with $\Omega_\nu$ greater than about 0.2 and that even a
small admixture of neutrinos has important consequences.

\medskip
\noindent{\bf Figure 6:}  $\Lambda$CDM models:
Acceptable values of $\Omega_\Lambda$
and $h$ for $n=0.9, 1.0$.  Note that large-scale structure
considerations generally favor a more aged Universe -- smaller
$h$ or larger $\Omega_\Lambda$.

\medskip
\noindent{\bf Figure 7:}  $\tau$CDM models:
Acceptable values of $g_*$ and $h$
for $n=0.9, 1.0$.  The quantity $g_*$ counts the total number
of relativistic species.  Photons and three massless neutrino species
correspond to $g_*=3.36$; for the equivalent of
$N_\nu$ massless neutrino species $g_* = 2. + 0.454N_\nu$.

\medskip
\noindent{\bf Figure 8:}  Isochrones in the $H_0$ - $\Omega_{\rm Matter}$
plane.  The green band corresponds to time back to the bang of
between $13\Gyr$ and $19\Gyr$; the yellow between $10\Gyr$ and $13\Gyr$;
red indicates $\Omega_{\rm Matter} < 0.3$ or expansion time less than
$10\Gyr$.  Broken horizon lines indicate the range
favored for the Hubble constant, $80\kms\Mpc^{-1} > H_0 >
60\kms\Mpc^{-1}$.  The age -- Hubble
constant tension is clear, especially for the inflationary prediction
of $\Omega_{\rm Matter} = 1$.  The broken curves denote the
$13\Gyr - 19\Gyr$ isochrone for $\Lambda$CDM; a cosmological constant
greatly lessens the tension.

\medskip
\noindent{\bf Figure 9:}  Predicted angular power spectra of
CBR anisotropy for several viable CDM models and the anticipated
uncertainty from a CBR satellite experiment with
angular resolution of $0.3^\circ$.  From top to bottom the models
are:  CDM with $h=0.35$, $\tau$CDM with the energy equivalent
of 12 massless neutrino species, $\Lambda$CDM with $h=0.65$ and
$\Omega_\Lambda = 0.6$, $\nu$CDM with $\Omega_\nu = 0.2$,
and CDM with $n=0.7$ (unspecified parameters have their standard
CDM values).

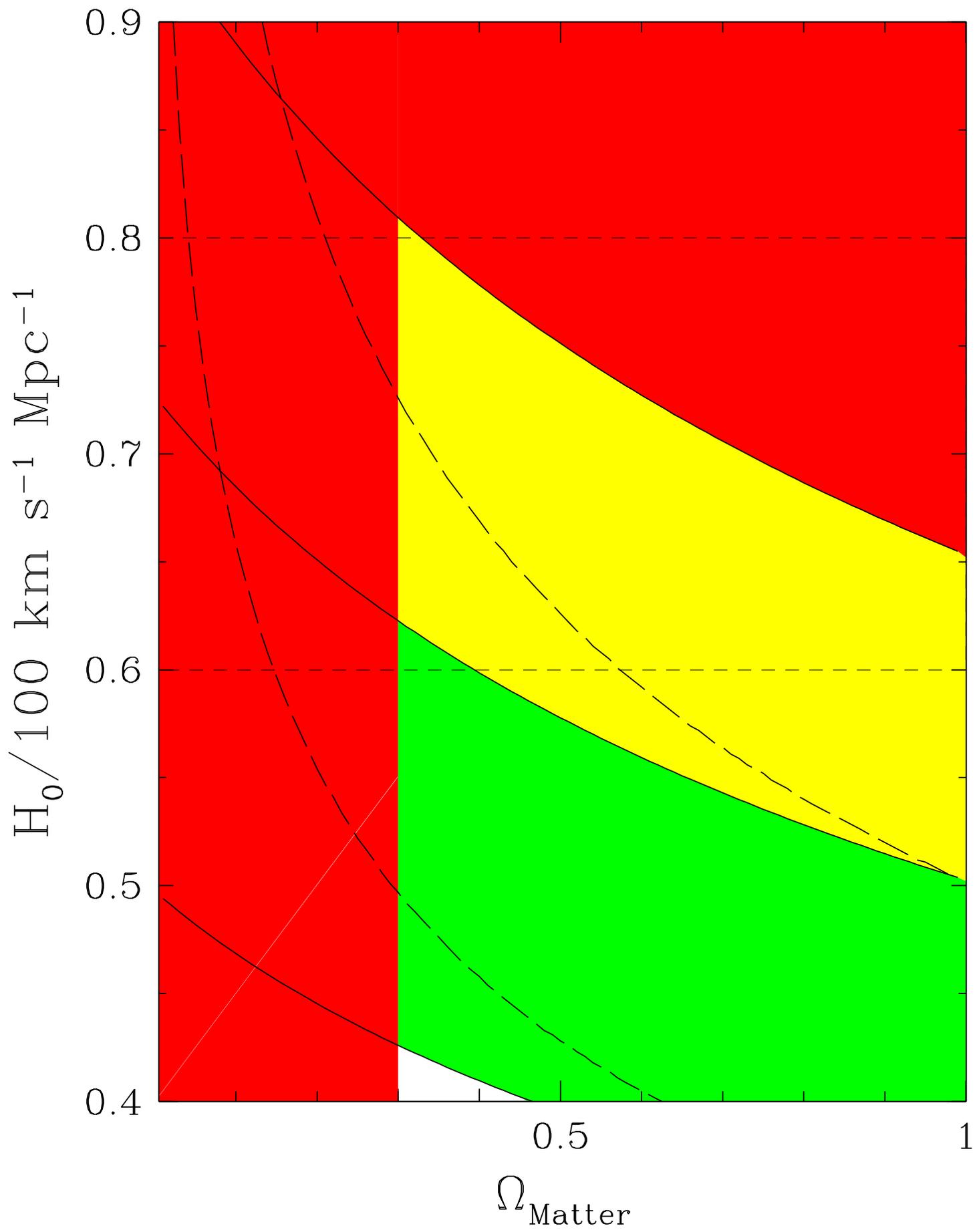
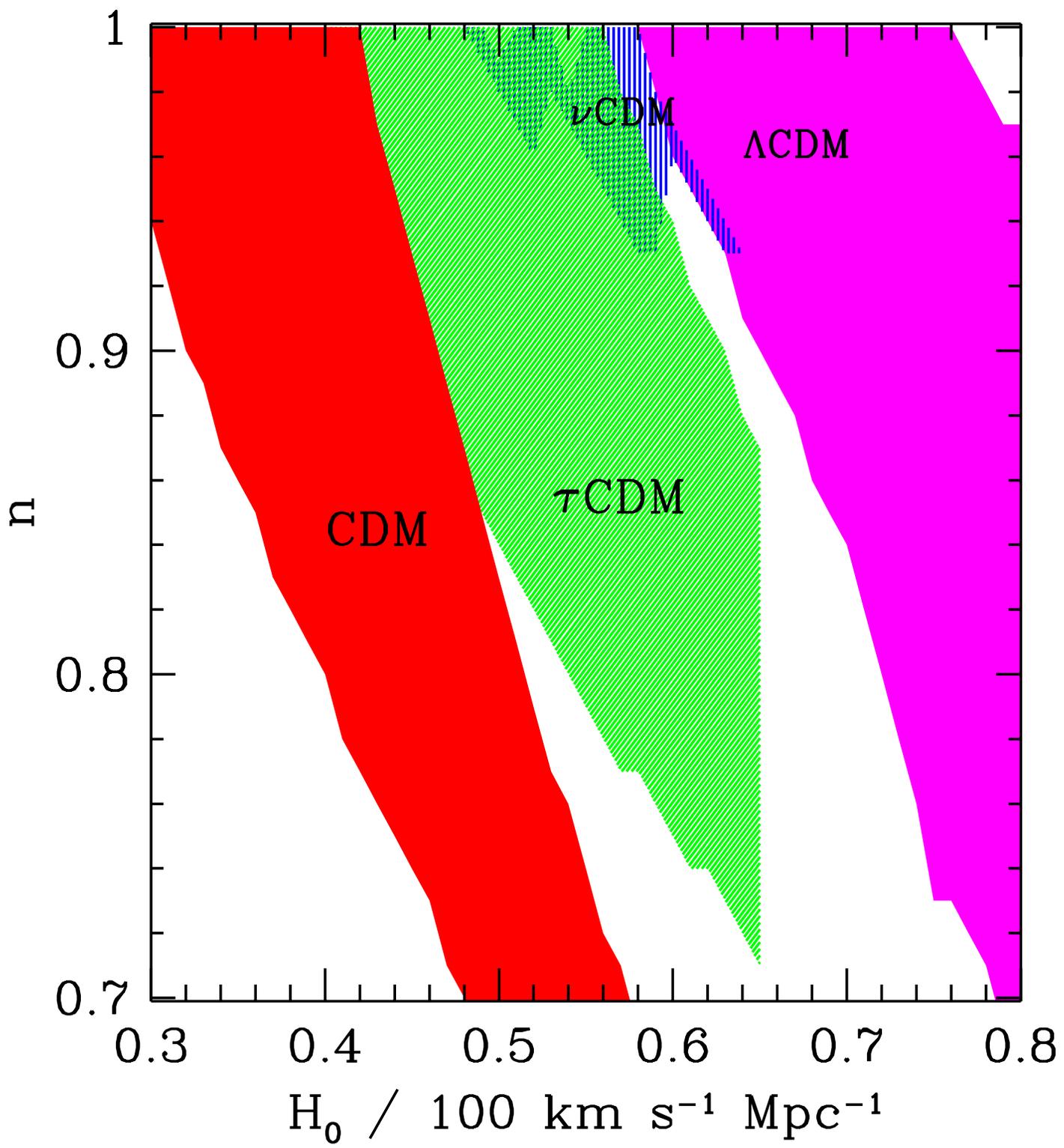
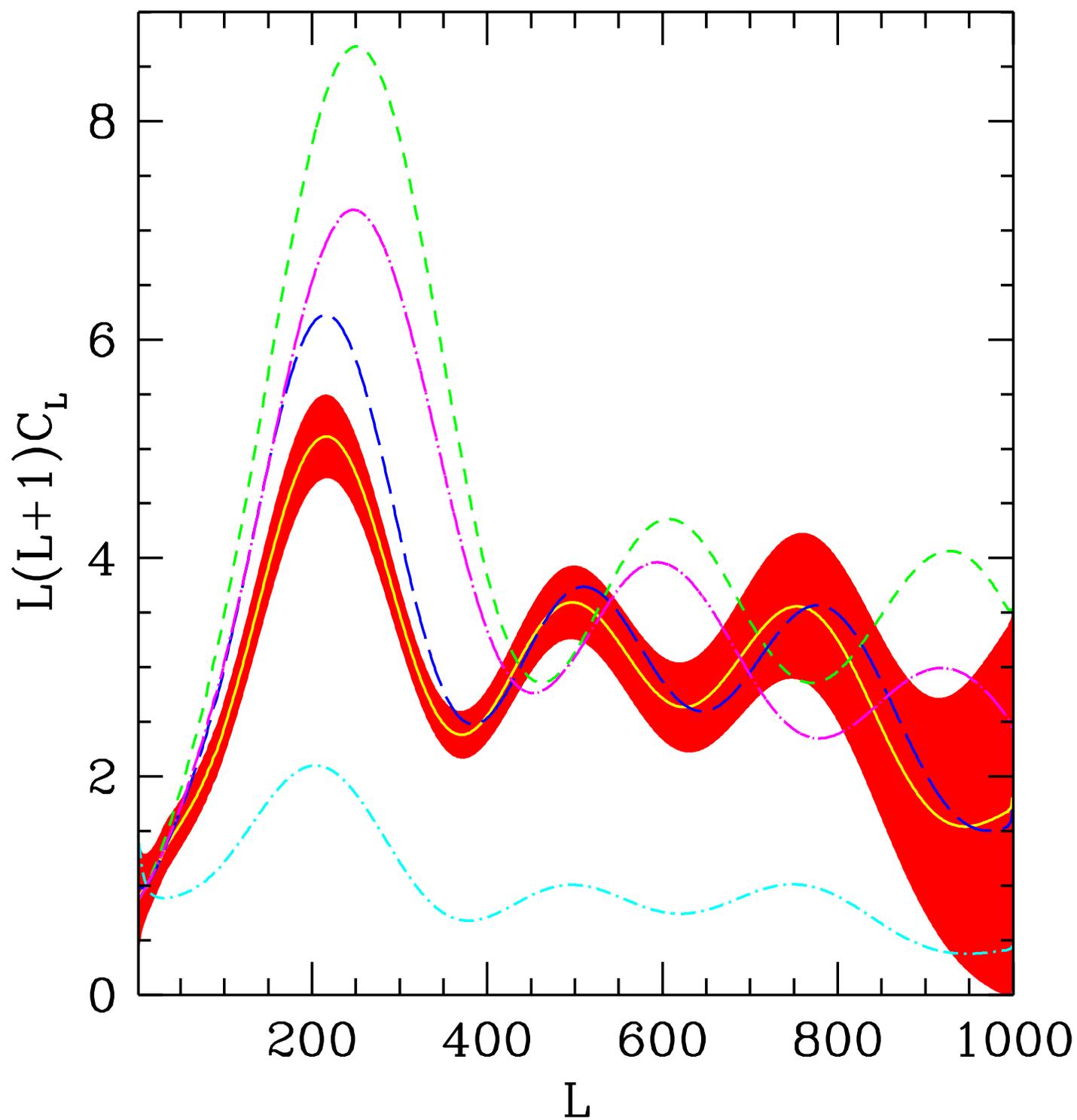
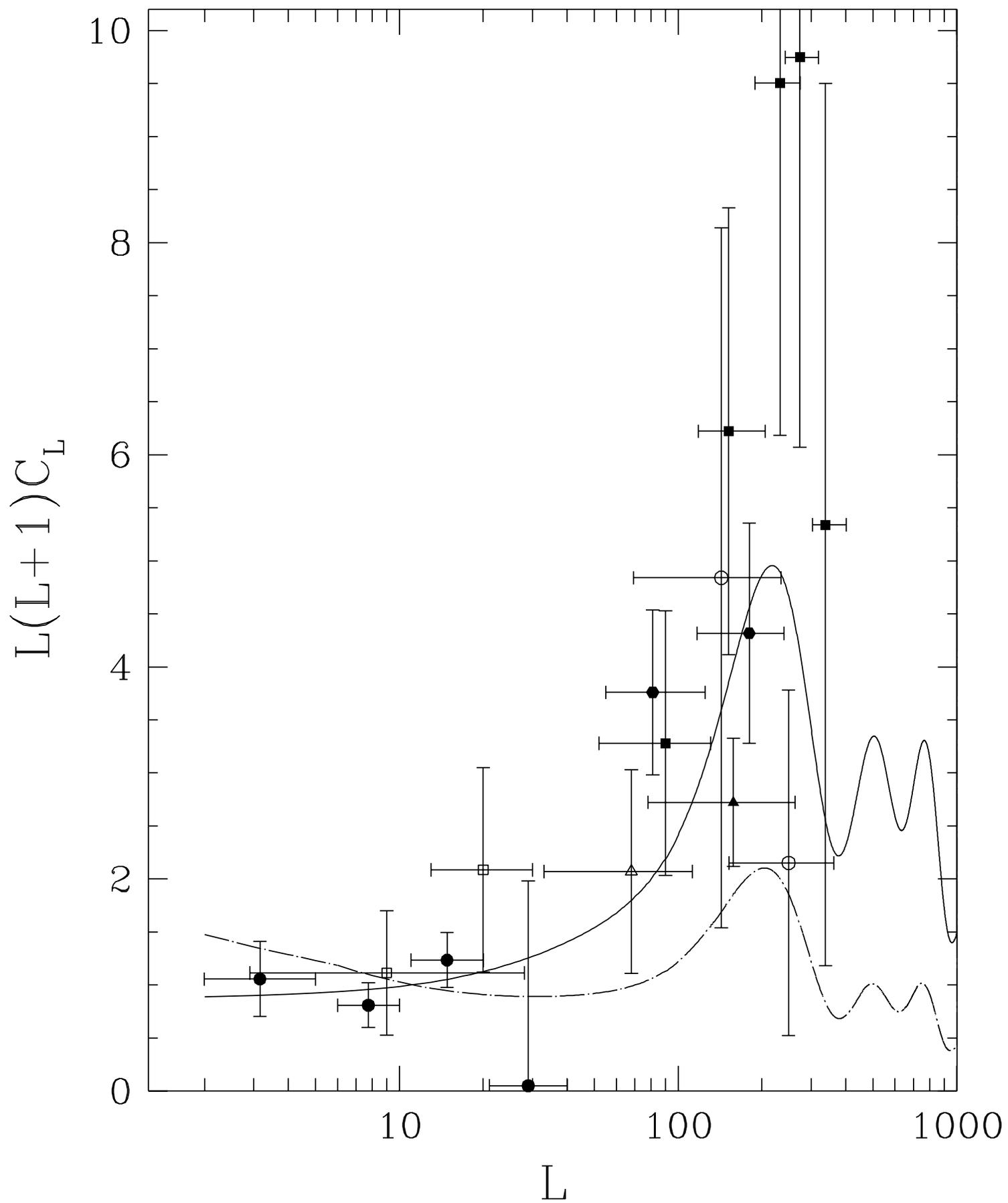
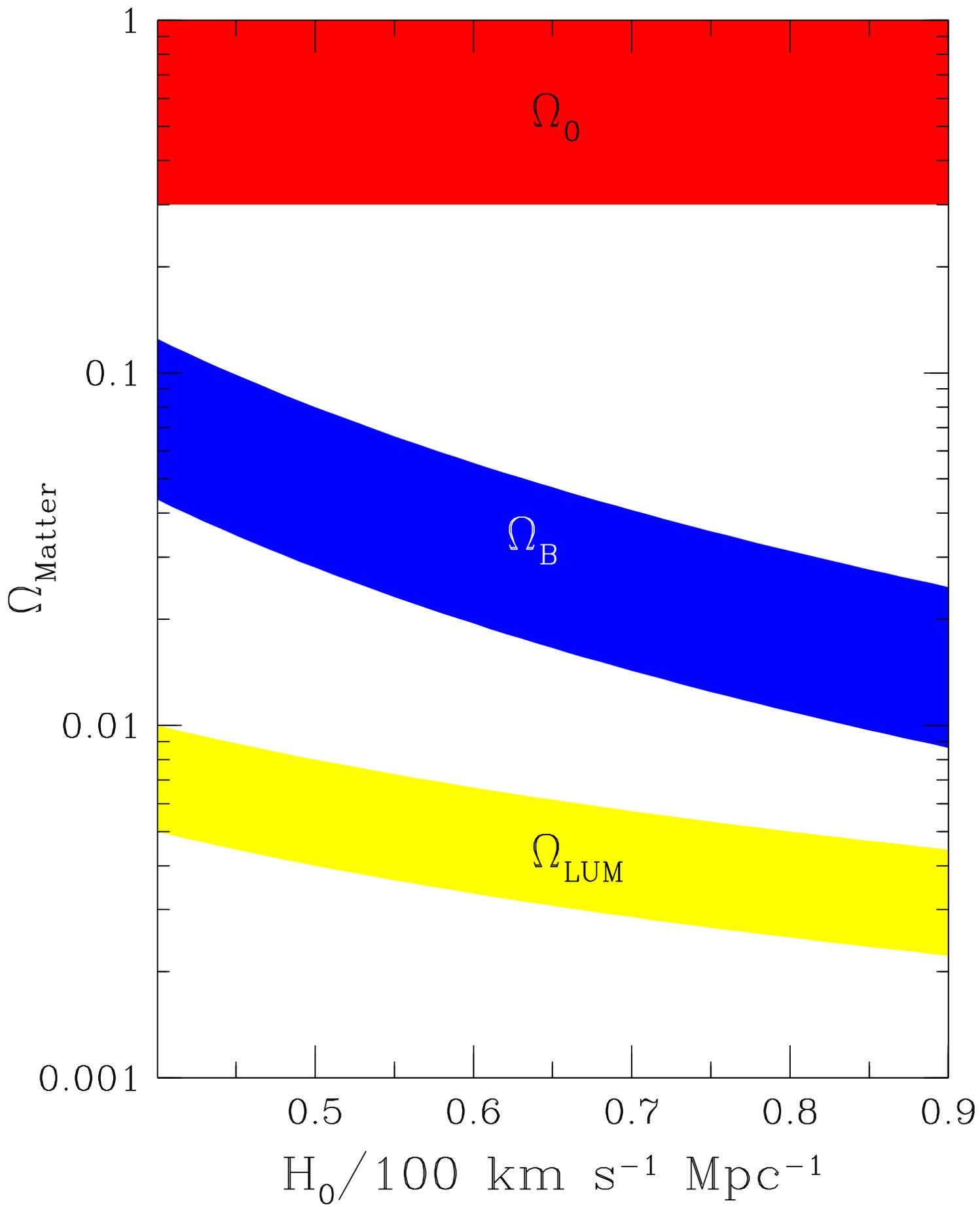
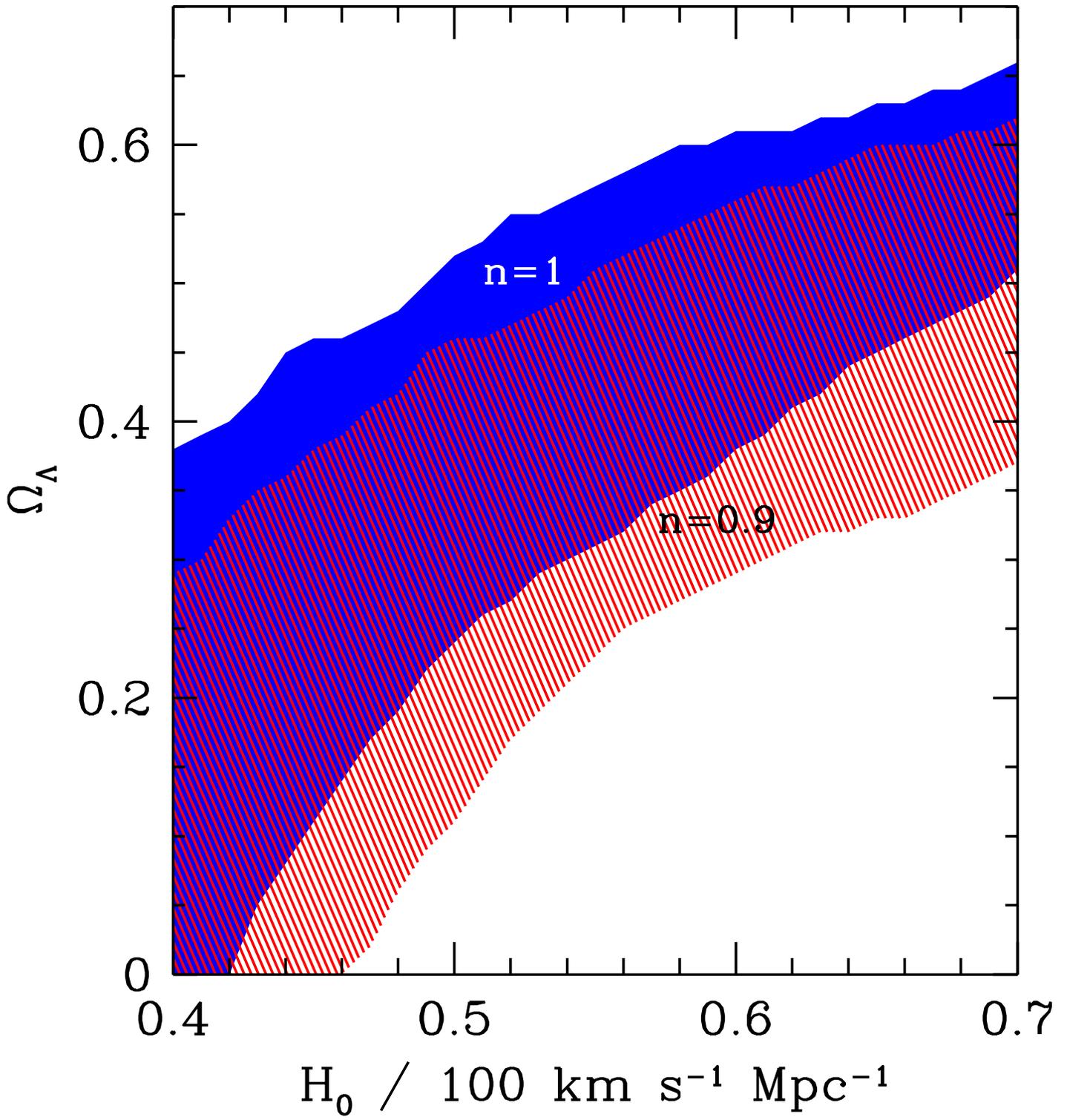
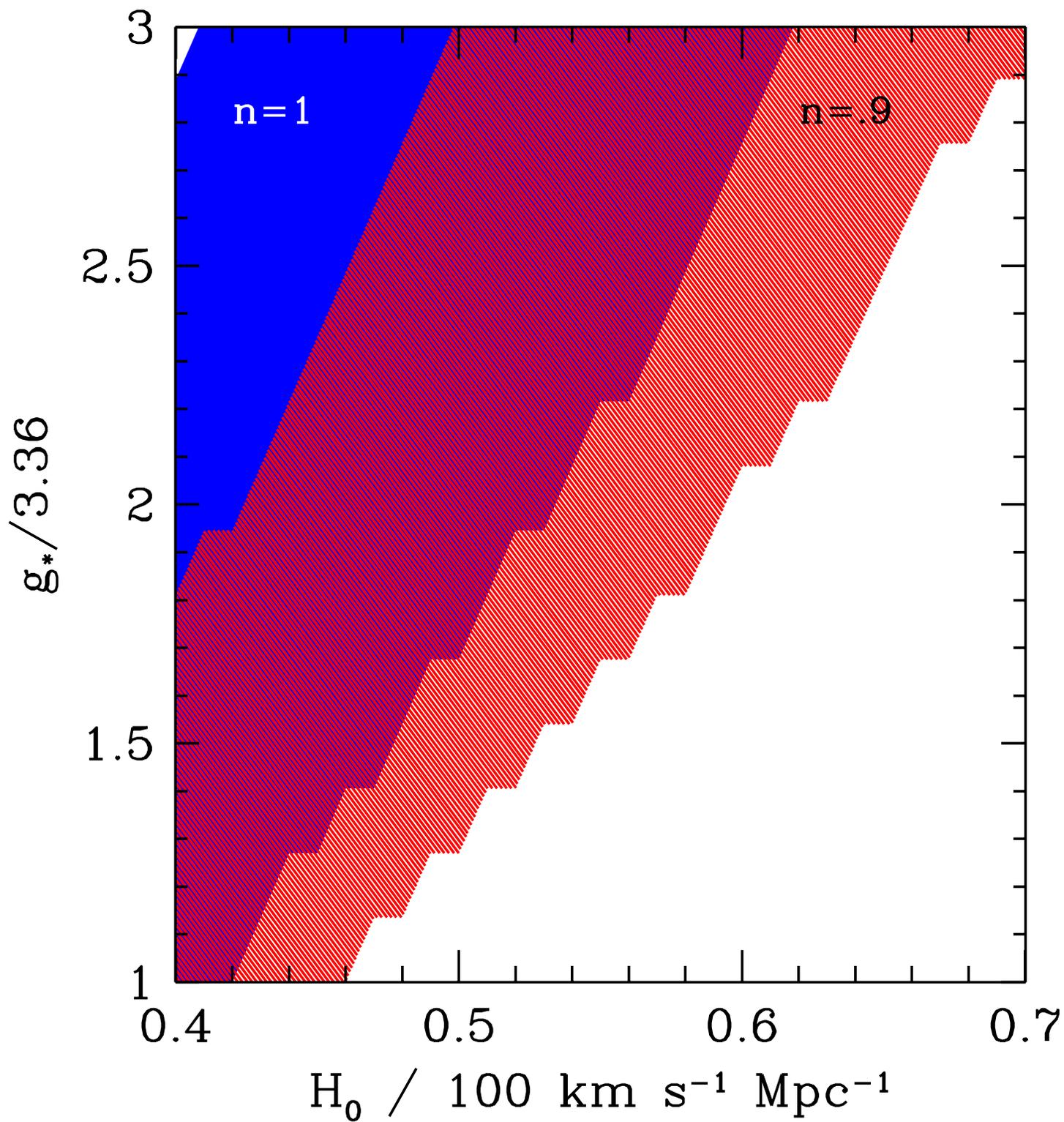
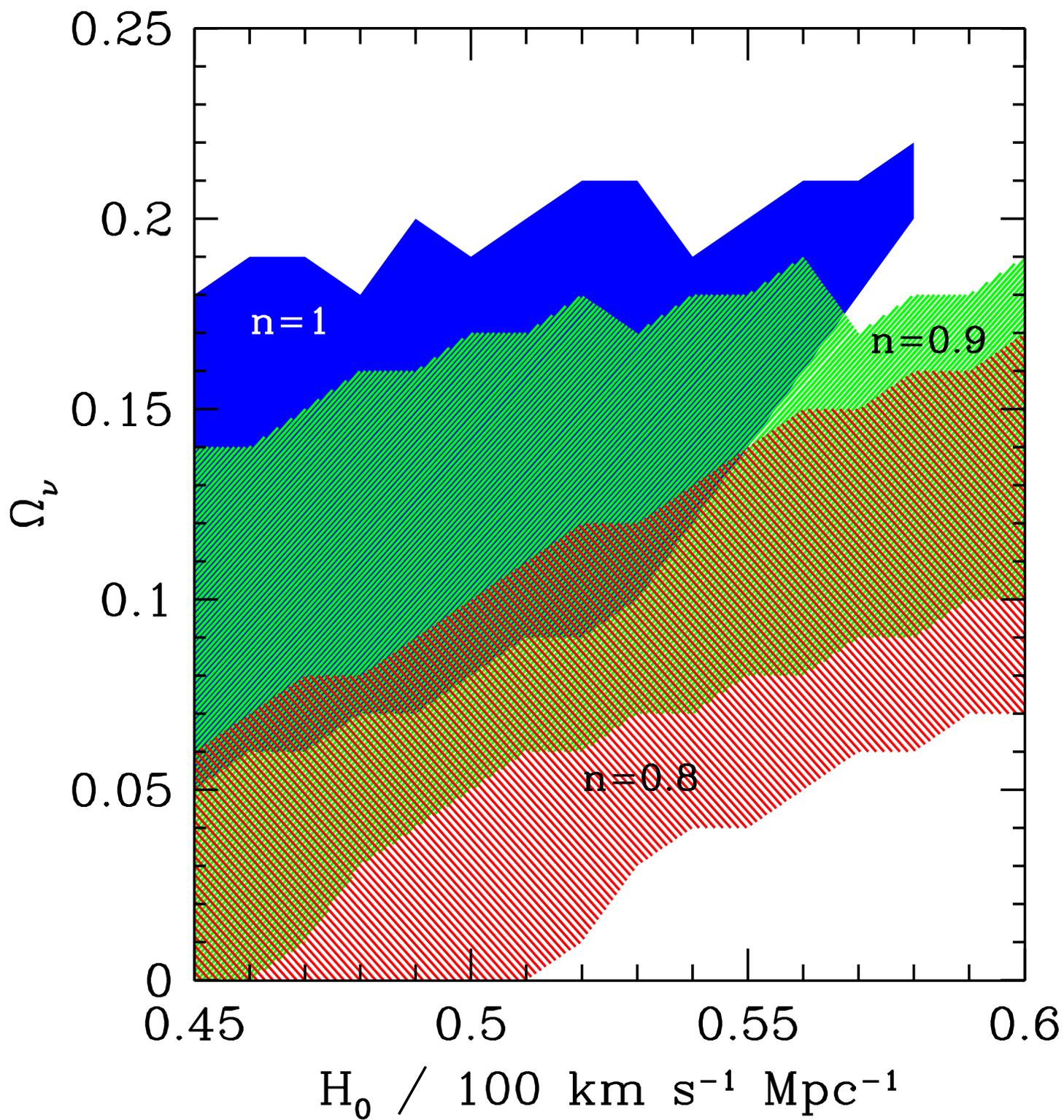
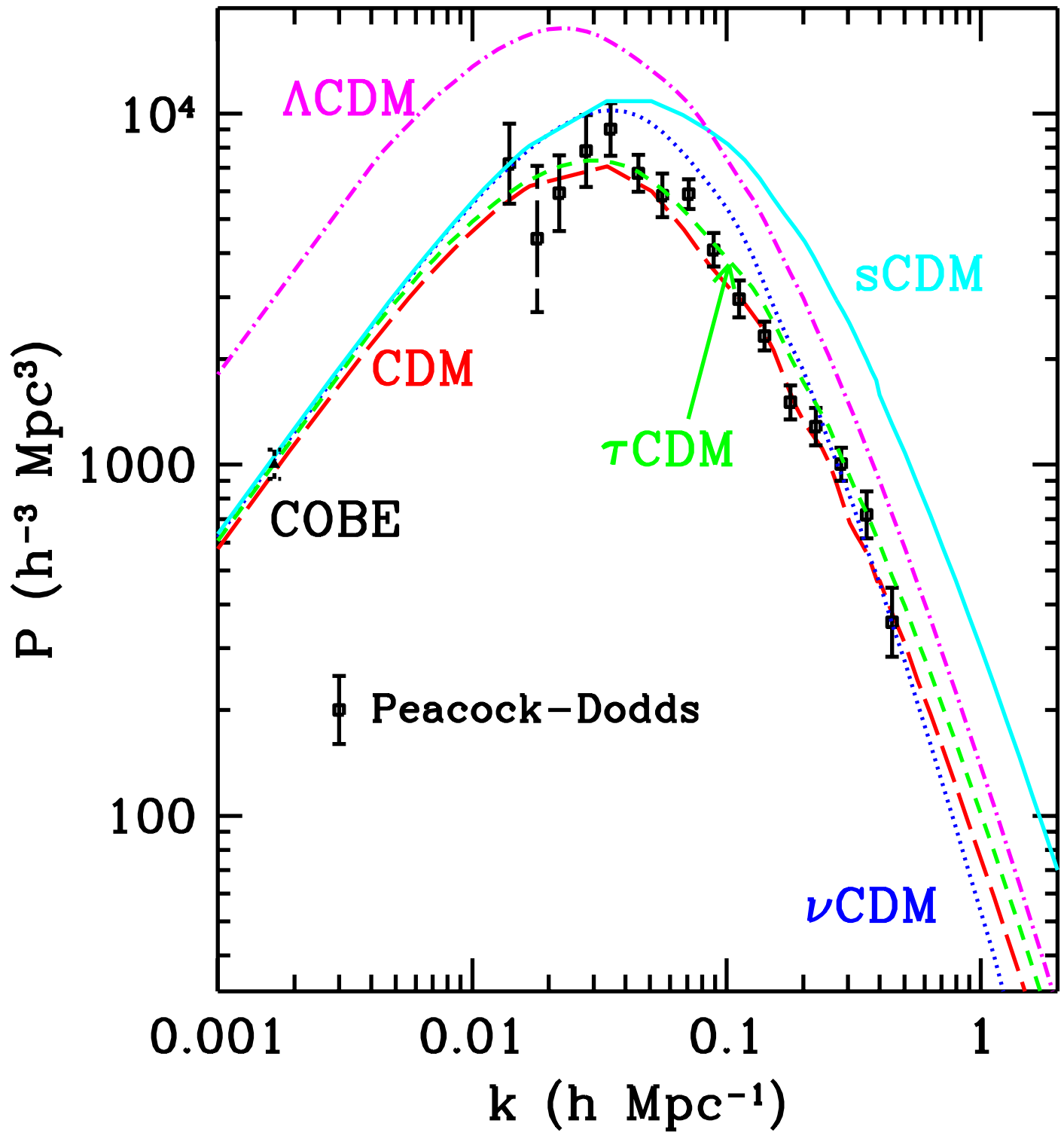
\end{document}